\documentclass[twocolumn]{aa} 
\usepackage{graphicx}
\usepackage{deluxetable}
\usepackage{natbib}
\usepackage{txfonts}
\begin{document}
   \title{Proper motions of cool and ultracool candidate members in the Upper Scorpius OB association}

   \author{H. Bouy\inst{1}\fnmsep\thanks{Marie Curie Outgoing International Fellow MOIF-CT-2005-8389}
          \and
           E.~L. Mart\'\i n\inst{2,3}
          }

   \offprints{H. Bouy}
   \authorrunning{Bouy \& Mart\'\i n}
   \titlerunning{Proper motions of Upper Scorpius cool and ultracool populations}

   \institute{Instituto de Astrof\'\i sica de Canarias, C/ V\'\i a L\'actea s/n, E-38200 - La Laguna, Tenerife, Spain \\
              \email{bouy@iac.es}
         \and
             University of Central Florida, Department of Physics, P.O. Box 162385, Orlando, FL 32816-2385, USA\\
        \and
             CAB-CSIC/INTA, Ctra de Torrejón a Ajalvir km 4, 28850 Torrejón de Ardoz, Madrid, Spain  \\              
}

   \date{Received ; accepted 01-07-2009}

 
  \abstract
  {}
   {Proper motion measurements of the cool and ultracool populations in the 
Upper Scorpius OB association are crucial to confirm membership and to identify possible run-away objects.}
   {We cross-match samples of photometrically selected and spectroscopically confirmed cool and ultracool  
(K5$<$SpT$<$M8.5) candidate members in the Upper Scorpius OB association using the literature and the USNO-B and the UCAC2 catalogues. 
251 of these objects have a USNO-B and/or UCAC2 counterpart with proper motion measurements.}
   {A significant fraction (19 objects, 7.6$\pm$1.8\%) of spectroscopically confirmed young objects show discrepant proper motion. They must either belong to unidentified coincident foreground associations, or originate from neighboring star forming regions or have recently experienced dynamical interactions within the association. The observed accretor and disc frequencies are lower among outliers, but with only 19 objects it is unreliable to draw firm statistical conclusions. Finally, we note that transverse velocities of very low mass members are indistinguishable from those of low mass members within $\approx$4~km s$^{-1}$.} 
   {}


   \maketitle
%

\section{Introduction}
The galaxy is made of a number of large scale stellar structures such as clusters, star forming regions, and OB associations. Kinematics has been successfully used to identify these associations \citep[e.g][]{2008arXiv0808.3362T, 2004ApJ...613L..65Z, 2006ApJ...649L.115Z}. The kinematic properties of the different classes of objects in a given association additionally hold important clues about its members history. Some models of formation predict that the early dynamical evolution of the parent proto-stellar cluster should lead to mass-dependent kinematic distributions \citep{2003MNRAS.346..369K} while other numerical simulations predict on the contrary similar kinematic properties over the whole mass spectrum \citep{2003MNRAS.339..577B,2009MNRAS.392..590B}. The kinematic study of low mass stars (LMSs, defined here as objects with spectral type {\bf M0$<$SpT$<$K5}), very low mass stars (VLMSs, M0$<$SpT$<$M5) and brown dwarfs (BDs, SpT$>$M6) in young associations offers a unique opportunity to test these predictions and the corresponding models of formation.

A number of surveys have used proper motions to identify and confirm new VLMS and BD members in young associations and clusters \citep[see e.g][]{2001A&A...367..211M, 2007MNRAS.378.1131C,2007AJ....134.2340K,2007AAS...210.9601C,2007MNRAS.380..712L,2007MNRAS.374..372L}, and various authors used radial velocity measurements to study the kinematic properties of young LMS, VLMS and BD \citep{2001A&A...379L...9J,2006A&A...448..655J,2006MNRAS.371L...6J, 2008MNRAS.385.2210M}.  In this paper, we make use of the USNO-B and UCAC2 catalogues of transverse velocity measurements to study the kinematic properties of LMS, VLMS and BD in the Upper Scorpius association.

\section{The USNO-B and UCAC2 catalogues}

The USNO-B catalogue \citep{2003AJ....125..984M} is made from a compilation of photographic plates taken from various sky surveys performed over the last 50 years. It provides  positions, proper motions and magnitudes in several photographic passbands over the entire sky. The catalogue claims to be complete up to V=21~mag, and to have an astrometric accuracy better than 0\farcs2 at J2000. It therefore provides a unique database of multi-epochs photometric and astrometric measurements. The oldest images were obtained as early as 1949, while the latest were obtained in 2002, offering a time-base as long as 53~yr in some cases. Uncertainties on the proper motion measurements are typically of the order of 1--5~mas yr$^{-1}$. 

The second US Naval Observatory CCD Astrograph Catalog \citep[UCAC2, ][]{2004AJ....127.3043Z} provides astrometry, photometry and proper motion measurements over a sky area between -90\degr\, to +40\degr\, declination, thus including Upper Scorpius. The survey was conducted between 1997 and 2004, and used a single bandpass intermediate between V and R (579--642 nm, hereafter UCmag) reaching a limiting magnitude UCmag$\approx$16~mag. The UCAC2 catalogue provides proper motion measurements based on the combination of the UCAC measurements and older catalogues. For stars fainter than V$\approx$12.5~mag, the Yellow Sky 3.0 catalogue was used as first epoch and proper motion measurements were made using these 2 epochs only. In the case of our study, uncertainties are typically of the order or 8--10~mas yr$^{-1}$, thus significantly larger than the USNO-B ones. All stars with high proper motions ($\ge$200 mas yr$^{-1}$) have been excluded from the UCAC2 catalogue \citep{2004AJ....127.3043Z}.

\section{The Upper Scorpius low and very low mass population and the USNO-B.1 and UCAC2 catalogues}
Over the last decade, Upper Scorpius has been one of the most targeted star forming regions for the search and study of LMS, VLMS and BD. In the present article, we combine samples from \citet{2000AJ....120..479A,2002AJ....124..404P,2004AJ....127..449M, 2008ApJ...688..377S} and \citet{Martin_inprep}. These five samples were selected in color-magnitude and color-color diagrams and confirmed with spectroscopy. Objects of these samples without spectroscopic confirmation were discarded. Other surveys discovered additional samples of Upper Scorpius low and very low mass members \citep[e.g ][]{2007MNRAS.374..372L} but used different selection criterion (photometry and kinematics) and did not include spectroscopic confirmation. For consistency we do not include them in our study.  Upper Scorpius was recently resolved in two distinct populations \citep{1996A&A...307..121B,2000A&A...356..541K,2007ApJ...662..413K}. The original surveys used in our study were all performed in Upper Scorpius~A (located east of 16$^{\rm h}$ and north of -28\degr following the definition of the above mentioned authors).

Table~\ref{surveys} gives an overview of the completeness limits of the five surveys. Even though a direct comparison is not possible since they used different photometric systems, Table~\ref{surveys} suggests that there is an overlap between the USNO-B catalogue (complete up to V=21~mag, corresponding to R$\sim$19~mag for an  M8 dwarf member in Upper Scorpius) and these surveys. Similarly, the shallower (UCmag$\lesssim$16~mag) UCAC2 catalogue is expected to be sensitive to Upper Scorpius members with spectral classes $\lesssim$M4.

The two catalogues provide relative\footnote{with respect to the YS4.0 catalog for USNO-B and YS3.0 catalog for UCAC2} proper motion measurements  and a direct comparison between the UCAC2 and USNO-B measurements is therefore not possible. The two catalogues have some remaining unknown systematic errors that can be significant for faint objects such as the ones in the combined sample. These errors are mostly related to the technology (a combination of CCD and Schmidt plates in the case of UCAC2 and Schmidt plates only in the case of USNO-B) and algorithms used to compute the proper motions. In the current state of the two catalogues, one cannot expect a very good agreement (Zacharias, private communication). We therefore chose to perform two independent analyses of the proper motion of the combined sample with each of the two catalogues. This approach offers the additonal advantage of providing an independent check for the objects in common in the two catalogues.

\subsection{USNO-B}
We cross-matched the lists of spectroscopically confirmed members found in the above mentioned surveys with the USNO-B.1 catalogue within a radius of 10\arcsec. Figure~\ref{distrib_r} shows the distribution of separation between the objects and their USNO-B closest match within 10\arcsec. It appears to be roughly normal at short distances with a peak around 0\farcs7 and a standard deviation of 0\farcs5. Allowing a large search radius greatly increases the probability of spurious matches, but limiting the search radius to a small value would prevent us from detecting real high proper motion objects. As a compromise, we limit the search to within 3\arcsec, corresponding to 4$\sim$5-$\sigma$ with respect to the peak in the separation distribution. 

 The vast majority of the objects (476 out of 515) have a counterpart for at least one of the 5 epochs of the USNO-B catalogue. The USNO-B catalogue being a compilation of several photographic surveys with distinct sensitivities and resolutions, not all the sources have been detected at several epochs and therefore only a fraction of them (227 out of 476) have proper motion measurements available. The other are in general too faint or too red to have been detected in the early surveys. Table~\ref{counterparts} gives an overview of the fraction of sources with proper motion measurements for the different surveys and for the combined sample.

\subsection{UCAC2}
\begin{figure}
   \centering
   \includegraphics[width=0.45\textwidth]{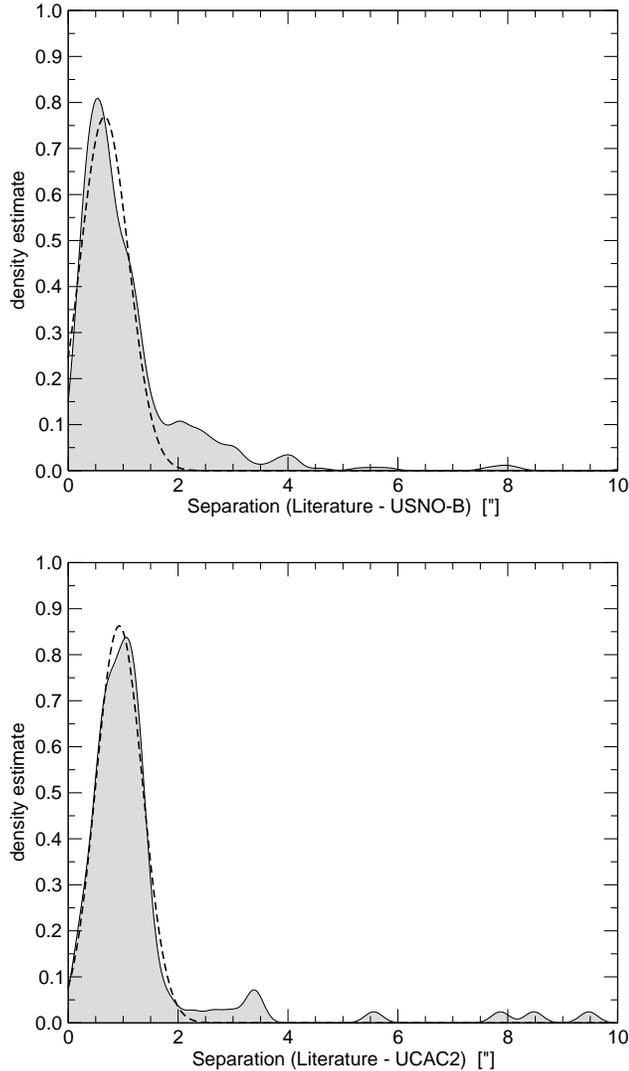}
   \caption{Distribution of separation between the closest USNO-B (top panel) or UCAC2 (bottom panel) match and our targets within 10\arcsec. The two distributions are roughly normal at short distances, and are well fitted by a Gaussian (dashed line) centered at 0\farcs7 and with $\sigma=$0\farcs5 in the case of USNO-B and centered at 0\farcs9 and with $\sigma$=0\farcs5 in the case of UCAC2. The distributions were computed using the kernel density estimation method with a Gaussian kernel and an optimized bandwidth of 0\farcs16 \citep{kernel_estimate_2d}.}
              \label{distrib_r}%
    \end{figure}

The same analysis was made for the UCAC2 catalogue. Figure~\ref{distrib_r} shows the distribution of separation between the objects and their UCAC2 closest match within 10\arcsec. As in the case of the USNO-B catalogue, the distribution is roughly normal at short distances, and peaks around 0\farcs9 with a standard deviation of 0\farcs5. We use the same 3\arcsec\, search radius, corresponding to 4$\sim$5-$\sigma$. A total of 90 objects among the 515 of the combined samples have a UCAC2 counterpart with a proper motion measurement within 3\arcsec. Table~\ref{counterparts} gives the fraction of sources with UCAC2 counterparts for each surveys and for the combined sample. The mean epoch of the UCAC2 CCD observations ranges between 1996.0 and 1999.8. A total of 66 objects of the combined sample have a counterpart within 3\arcsec\, and with a proper motion measurement in both the USNO-B and UCAC2 catalogues. 

\begin{center}
\begin{table}
\caption{Completeness limits and epochs of the different surveys}
\label{surveys}
\begin{tabular}{lccc}\hline\hline
Reference                      & R             & I           & Epoch Range  \\
                               & [mag]         & [mag]       & [yr]         \\
\hline                                                                       
\citet{2002AJ....124..404P}    & 16.5          & \nodata     & 2000.4-2001.4 \\
\citet{2000AJ....120..479A}    & 19.0          & 18.5        & 1998.2-1991.4 \\
\citet{2004AJ....127..449M}    & \nodata       & 18.0        & 1996.2-2000.1 \\
\citet{Martin_inprep}          & \nodata       & 18.0        & 1996.2-2000.1 \\
\citet{2008ApJ...688..377S}    & 19.0          & 18.0        & 2004.4        \\
\hline
\end{tabular}

Note -- The surveys used different R and I-band photometric systems. \citet{2008ApJ...688..377S} survey used $r$ and $i$ rather than R and I. Refer to the corresponding articles for more details on each photometric system.
\end{table}
\end{center}

\begin{center}
\begin{table*}
\caption{Number of objects with a USNO-B and UCAC2 counterpart}
\begin{scriptsize}
\label{counterparts}
\begin{tabular}{lcccc}\hline\hline
Survey                         & Number of     & Total USNO-B & USNO-B counterparts  & UCAC2          \\
                               & objects       & counterparts & with proper motion   & counterparts   \\
\hline                                                                       
\citet{2002AJ....124..404P}    & 166           & 166          & 123                  & 87            \\
\citet{2000AJ....120..479A}    & 20            & 20           & 12                   & 1             \\
\citet{2004AJ....127..449M}    & 28            & 23           & 14                   & 0             \\
\citet{Martin_inprep}          & 7             & 6            & 3                    & 0             \\
\citet{2008ApJ...688..377S}    & 127           & 121          & 93                   & 2             \\
\hline 
Total                          & 515           & 476          & 227                  & 90             \\
\hline
\end{tabular}

Note -- All UCAC2 sources have a proper motion measurement.
\end{scriptsize}
\end{table*}
\end{center}

\section{Reliability of the cross-identification \label{cross-match}}
Because of their size, all-sky catalogues necessarily contains errors and problems. In the following section, we refine the cross-matching analysis in order to reject problematic sources.

\begin{figure}
   \centering
   \includegraphics[width=0.45\textwidth]{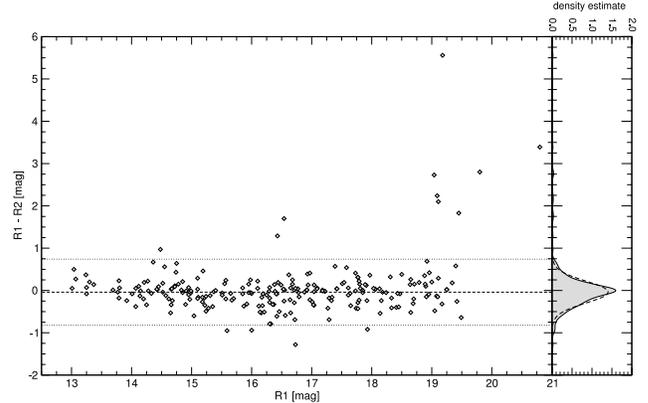}
   \caption{Difference between the R1 and R2 magnitudes as a function of the R1 magnitude for the USNO-B matches of the combined sample. The density estimate is also represented (grey). It is roughly normal (dashed curve), centered around -0.04~mag and with a width of $\sigma$=0.26~mag.  Any USNO-B source outside the $\pm$3-$\sigma$ limits was discarded.}
              \label{phot_test}%
    \end{figure}

In the current release of the USNO-B catalogue, the combination of five independent photographic surveys into a single catalogue was based only on astrometric coincidence with no reference to the photometry, and the final catalogue is known to contain a number of spurious merges with erroneous proper motion measurements. \citet{2003AJ....125..984M} recommend to use the multi-epoch photometry to identify erroneous combinations in the catalogue. The catalogue provides photometric measurements in 5 photographic passbands: the 2 epoch blue magnitudes B1 and B2 (400--500~nm), the 2 epoch red magnitudes R1 and R2 (600--750~nm), and one epoch infrared magnitude I (750--1000~nm). The cool objects in our samples are detected in the reddest of these passbands (R1, R2 and I) and we check the consistency of the 2 epoch R1 and R2 to ensure that each USNO-B entry corresponds to a valid source. Figure~\ref{phot_test} shows the difference (R1-R2) as a function of R1. The distribution is normal, centered on (R1-R2)=-0.04~mag and with $\sigma$=0.26~mag. A total of 14 sources with a second epoch R2 magnitude inconsistent with the first epoch R1 magnitude within 3-$\sigma$ were flagged as suspicious and discarded for the rest of the study. We also reject the one USNO-B counterpart flagged as an object on a diffraction spike.

The UCAC2 catalogue was built using only 2 epochs and 2 different passbands: the UCmag (intermediate between conventional V and R filters) in one case and the yellow Schmidt plates of the YS3 survey (similar to V-band) in the other case. It is therefore not possible to check the consistency of the photometry of the sources as in the case of USNO-B.

\section{Membership \label{membership}}

\begin{figure}
   \centering
   \includegraphics[width=0.45\textwidth]{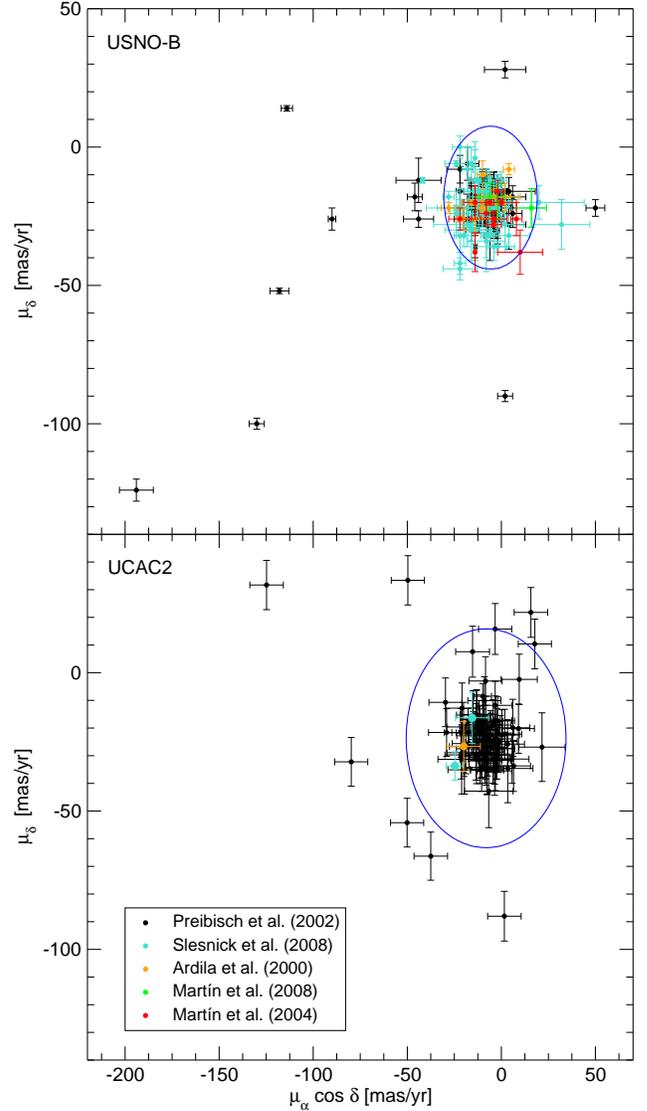}
   \caption{Vector point diagrams for the spectroscopically confirmed members of Upper Scorpius with USNO-B (upper panel) and UCAC2 (lower panel) counterparts. Blue ellipses represent the $\mathcal{P}=$0.1\% membership probability level used to select outliers candidates. \label{vectorpointdiag}}
    \end{figure}

Figure~\ref{vectorpointdiag} shows the vector point diagrams obtained with USNO-B and UCAC2 proper motion measurements for the combined sample. A clustering appears clearly around ($\mu_{\alpha} \cos{\delta}, \mu_{\delta}$) = (-6.7,-19.3)~mas yr$^{-1}$ (USNO-B) and ($\mu_{\alpha} \cos{\delta}, \mu_{\delta}$) = (-9.1,-24.7)~mas yr$^{-1}$ (UCAC2) and we associate it with the Upper Scorpius co-moving group. A significant number of objects fall far away from the co-moving group.

Proper motion can be used to quantitatively assess the membership of objects in a co-moving group. As described in \citet{1971A&A....14..226S}, the distribution of proper motion in a vector point diagram can be modeled by two independent overlapping distributions of member and interloper stars. In this model, the proper motion dispersion of members is assumed to be caused by observational and measurement errors normally distributed, so that the distribution of member stars is represented by a bivariate normal distribution. Unlike in the cases described by \citet{1971A&A....14..226S}, all the objects of our sample have been selected based on their spectroscopic characteristics, so that the contribution of non-members to the vector point diagram is mainly due to coincident young sources. We consider that their distribution in the vector point diagram is constant over the parameter space probed by our analysis and model it with a uniform function. A 2-D Gaussian + constant function is thus fitted to the 2-D kernel density estimate of the vector point diagram and we obtain:

\begin{eqnarray*}
f_{m\,\rm USNO}(\mu_{\alpha} \cos{\delta},\mu_{\delta}) & = & 0.0061 \exp \Bigg\{- \frac{1}{2}\Biggl[\Bigg(\frac{\mu_{\alpha} \cos{\delta} + 6.7}{5.0}\Bigg)^2 \\
 & + & \Bigg(\frac{\mu_{\delta}+19.3}{5.2}\Bigg)^2\Biggl]\Bigg\} \\
f_{m\,\rm UCAC2}(\mu_{\alpha} \cos{\delta},\mu_{\delta}) & = & 0.0022 \exp \Bigg\{- \frac{1}{2}\Biggl[\Bigg(\frac{\mu_{\alpha} \cos{\delta} + 9.1}{8.7}\Bigg)^2 \\
 & + & \Bigg(\frac{\mu_{\delta}+24.7}{8.1}\Bigg)^2\Biggl]\Bigg\} 
\end{eqnarray*}
and
\begin{eqnarray*}
f_{i\,\rm USNO}(\mu_{\alpha} \cos{\delta},\mu_{\delta}) = 2.9\times10^{-6} \\
f_{i\,\rm UCAC2}(\mu_{\alpha} \cos{\delta},\mu_{\delta}) = 1.5\times10^{-6}
\end{eqnarray*}

where $f_{m}(\mu_{\alpha} \cos{\delta},\mu_{\delta})$ and $f_{i}(\mu_{\alpha} \cos{\delta},\mu_{\delta})$ are the vector point distributions of respectively the members and interlopers. The membership probability can then be computed as:
\begin{displaymath}
\mathcal{P}=\frac{f_{m}(\mu_{\alpha} \cos{\delta},\mu_{\delta})}{f_{m}(\mu_{\alpha} \cos{\delta},\mu_{\delta}) + f_{i}(\mu_{\alpha} \cos{\delta},\mu_{\delta})}
\end{displaymath}

For the rest of the discussion we classify as highly probable members objects that are located inside the $\mathcal{P}=$0.1\% (3-$\sigma$) membership probability ellipse within the uncertainties, and as candidate outliers objects outside the $\mathcal{P}=$0.1\% membership probability ellipse. Even though this arbitrary threshold probably misses a number of outliers, it offers the advantage of separating obvious outliers from the overall co-moving group. The main goal for the rest of this paper is to confirm, identify and characterize the properties of these obvious outliers. The task of drawing firmer statistical conclusions regarding their frequency is left to more detailed studies of complete samples using higher proper motion accuracy. A total of 16 outliers are found in the USNO-B analysis and 9 in the UCAC2 analysis. We note that 3 objects (UScoCTIO-101, SCHJ16162599-21122315 and SCHJ16211564-24361173) have a probability membership less than 30\%. We classify them as likely outliers, but do not include them in the rest of our analysis for consistency.

\section{Reality of the outliers}

Because of the sample size, a human inspection of each USNO-B and UCAC2 counterpart is not possible. The rejection of doubtful objects described in section~\ref{cross-match} is purely statistical. Problematic sources might remain in the final sample. Outliers are of course suspicious in nature and deserve a more careful inspection before any further discussion can be made.

\subsection{Overlap between USNO-B and UCAC2} 
A total of 66 objects of the combined sample have a counterpart in both the USNO-B and UCAC2 catalogues (see Table~\ref{table_sample}). As mentioned above, the systematic errors of each catalogues prevent us from making a direct comparison between the USNO-B and UCAC2 proper motion measurements. On the other hand, the consistency of the membership probability for the 66 sources in common provides an important check of the reliability of the two analyses. Table~\ref{table_sample} shows that the membership probabilities derived independently using USNO-B and UCAC2 indeed agree well within $\pm$2\% for all 66 objects in common except for one (USco-161437.52-185824.0). With a membership probability of 80\% in USNO-B, and 99.2\% in UCAC2, we nevertheless classify this latter source as a highly probable member. Moreover, four of the 16 outliers found in the USNO-B analysis have a UCAC2 counterpart. All four are independently classified as outliers by the UCAC2 analysis as well. The two surveys being based on completely independent datasets, this comparison demonstrates the reliability and robustness of our membership analysis and classification criterion. 

\subsection{Probability estimator (USNO-B)}
Proper motions in the USNO-B catalogue are given with their uncertainties as well as a probability estimator for the likelihood that the proper motion is correct. This estimator ranges between 0.1 and 0.9, a larger value corresponding to a higher probability of being correct. It is a measurement of the goodness of fit between the observed motion over N$_{\rm obs}$ epochs reported in the catalogue and the linear motion fit. It is computed from the integral probability that the sample of N$_{\rm obs}$ random observations would yield an experimental linear-correlation as large or larger than the observed one \citep[as described in ][ D. Monet, private communication]{1992drea.book.....B}. 

For equal values of the probability estimator, sources with N$_{\rm obs}$=3 have a less believable proper motion measurements than sources with N$_{\rm obs}\ge$4. The probability estimator  is therefore not enough in itself to assess the reliability of the proper motion measurements. We note that at least 3 epochs have been used for all the USNO-B counterparts in our analysis. In fact, 69\% of the sources in our study have been reported 5 times out of 5 epochs, and 26\% of them have been reported 4 times out of 5 epochs (see Table~\ref{table_sample}) while only 5\% have been reported 3 times only out of 5 epochs. Only 1 out of the 16 outliers identified in the USNO-B survey has been reported 3 times only out of 5 epochs (SCHJ16162599-21122315). The probability estimator are therefore believed to be reliable for the 15 others. All 15 have a probability estimator of 0.9 estimated over 4 epochs or more (see Table~\ref{table_sample}).

\subsection{Visual inspection}
False detections as well as multiple systems are the main source of error for the USNO-B proper motion measurements (D. Monet, private communication). The USNO-B Schmidt plate 5 epoch images are available on-line. We retrieved the data for all 16 outliers identified in the USNO-B analysis as well as the corresponding 2MASS \citep{2006AJ....131.1163S}, DENIS \citep{2000A&AS..141..313F} and UKIDSS \citep{UKIDSS} images of both USNO-B and UCAC2 outliers when available to check the quality of the multi-epoch detections and the eventual presence of blended companions. Two objects are resolved as multiple systems and are discarded for the rest of the analysis. 

\medskip

\noindent \emph{USco-160726.8-185521} is resolved as a binary in \citet{2006A&A...451..177B} adaptive optics images with a separation of 3\farcs191, a position angle of 353.1\degr\, and a difference of magnitude in Ks of $\Delta m$=2.0~mag. It is also clearly elongated in the 2MASS images. The UCAC2 proper motion measurement is therefore suspicious and this object is removed from the list of outlier candidates.
\medskip

\noindent \emph{SCH~J16202523-23160347} is elongated or resolved as a close blended visual binary in most USNO-B images. It is also elongated in the 2MASS images and detected but unresolved in the DENIS images. The source is clearly resolved in the UKIDSS images as a visual triple system. This interesting object is discussed in more detail in Appendix~A, and is removed from the list of outlier candidates for the rest of the discussion as the presence of visual companions might have affected the USNO-B proper motion measurement. 

A total of 19 outliers {\bf (15 identified in USNO-B and 4 in UCAC2)} remain after the visual inspection.

\section{Kinematic properties of the Upper Scorpius population}

\begin{figure*}
   \centering
   \includegraphics[width=0.95\textwidth]{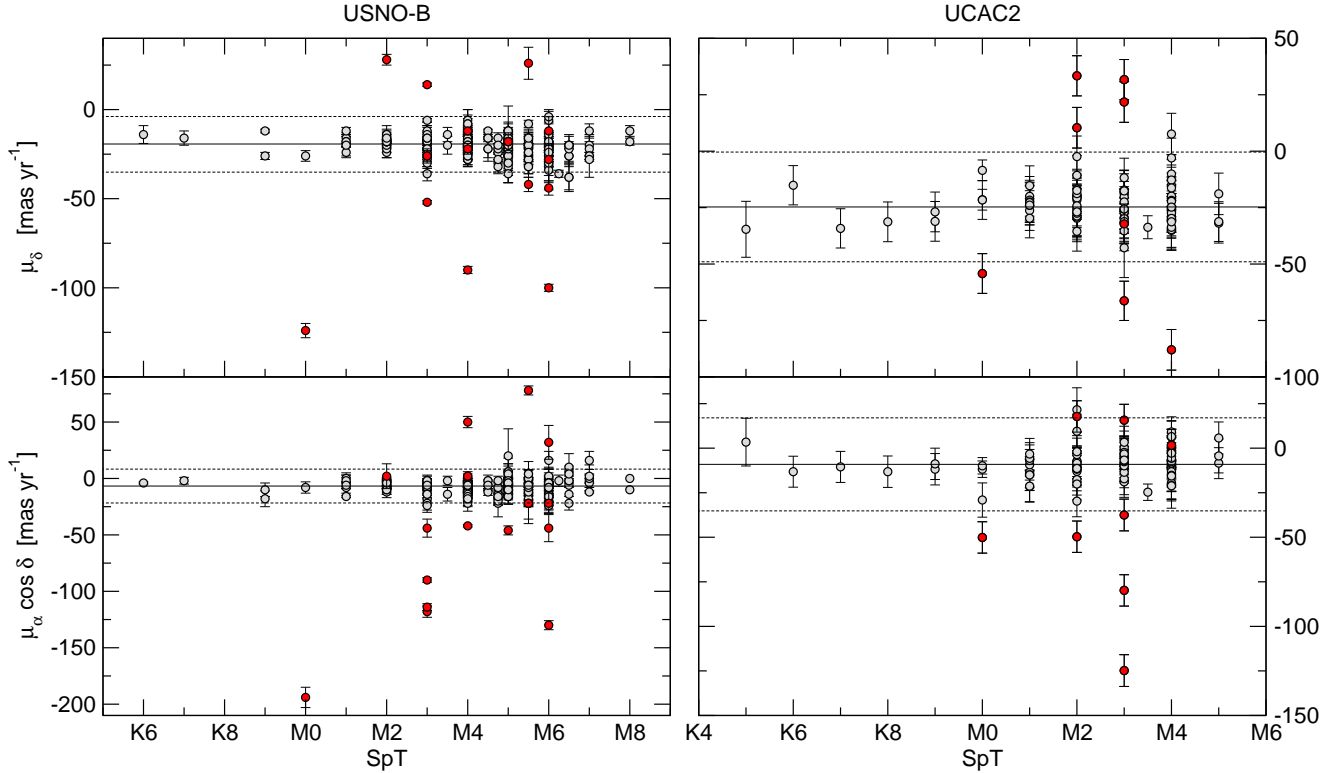}
   \caption{Transverse velocities as a function of spectral type for the USNO-B (left) and UCAC2 (right) analysis. Co-moving members are represented with grey dots, and outliers with red dots. The mean velocities (solid line) and the $\pm$3-$\sigma$ dispersion (dashed lines) of the association computed in section~\ref{membership} are also represented. There is no difference between early and late spectral types members within the individual uncertainties and overall dispersion. }
              \label{mu_spt}%
    \end{figure*}

Figure~\ref{mu_spt} shows the distribution of transverse velocities as a function of spectral type. No significant dependence on the spectral type can be seen for co-moving members within the relatively large individual uncertainties and overall dispersion. These uncertainties ($\approx$6~mas yr$^{-1}$ in average for USNO-B, corresponding to $\approx$4~km s$^{-1}$ at the distance of Upper Scorpius) are larger than the difference in dispersion velocities between stars and VLM objects expected from dynamical interactions in the early stages of young proto-stellar clusters \citep[$\approx$2~km s$^{-1}$,][]{2003MNRAS.346..369K,2003A&A...400.1031S,2009MNRAS.392..590B}. More accurate measurements are therefore required to improve this upper limit and make a detailed comparison of the kinematics of stellar and substellar members. Finally, we note that the spectral type distributions among outliers and members are indistinguishable at the 83\% confidence level as established by a Kolmogorov-Smirnov test.

\section{Nature of the contaminants \label{nature}}

\begin{figure*}
   \centering
   \includegraphics[width=0.95\textwidth]{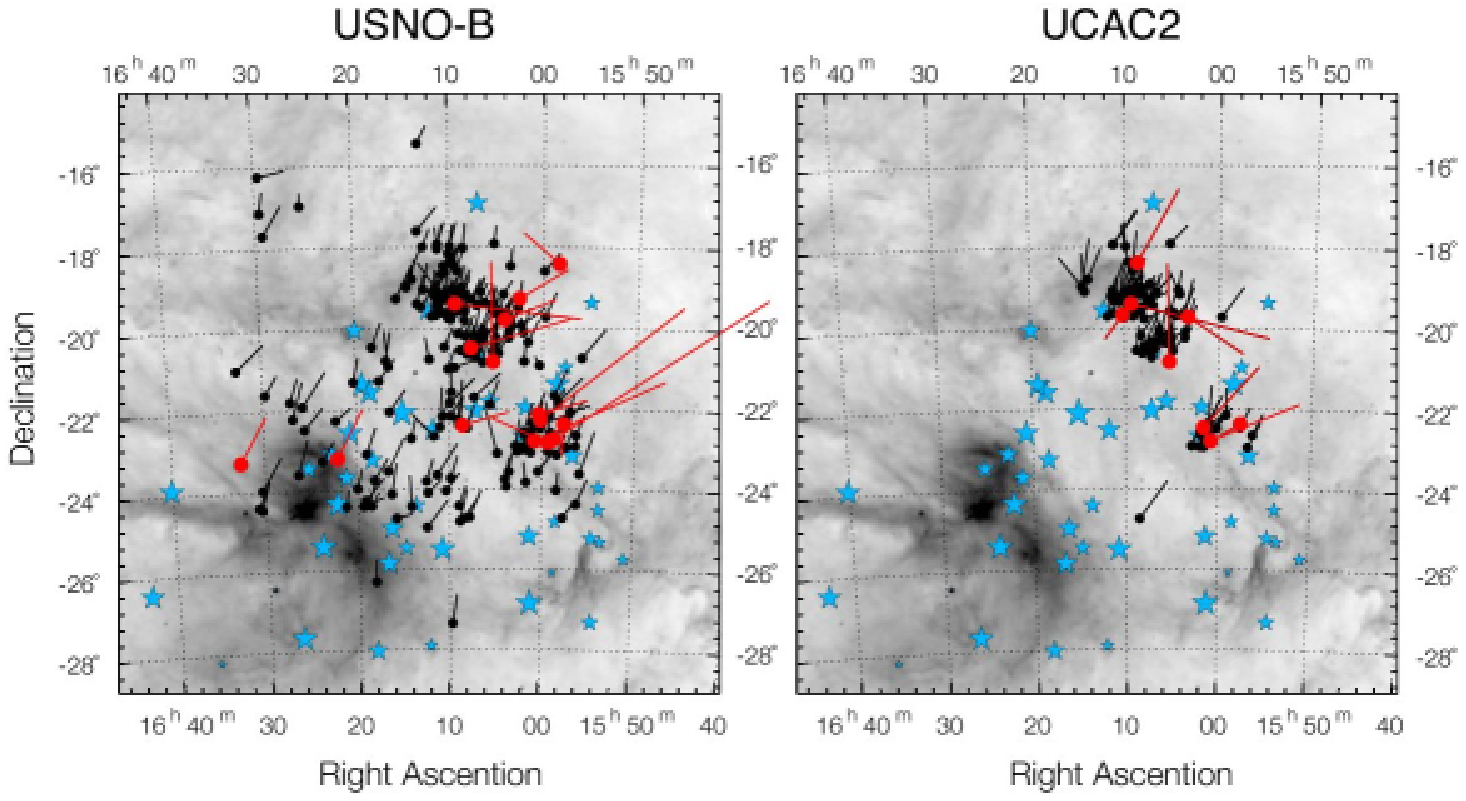}
   \caption{Spatial distribution of LMSs, VLMSs and BDs in our sample over-plotted on an IRAS 12~$\mu$m image for the USNO-B analysis (left) and UCAC2 analysis (right). Spectroscopically confirmed members with proper motion consistent with the co-moving group are represented by black dots. Candidate outliers are represented with red dots. Hipparcos members from \citet{1999AJ....117..354D} are represented with blue stars with sizes proportional to their V-band luminosities. The segments represent the objects motions over the past 100\,000~yr, assuming constant proper motions and null radial velocities. The densest nebulosity in the lower left quadrant are part of the $\rho-$Oph molecular cloud.}
              \label{map}%
    \end{figure*}

All the objects in the initial sample were confirmed as members of the association based on spectroscopic characteristics unambiguously associated with youth (atomic and/or molecular age diagnostics such as the presence and strength of Li and \ion{Na}{I} absorption, or H$\alpha$ emission). These different diagnostics correspond to different limits on the age of these objects. \citet{2002AJ....124..404P} Li selection criterion places an upper limit on the age at $\lesssim$30~Myr. \citet{2008ApJ...688..377S} \ion{Na}{I} criterion \citep[which is similar to that of][]{2004AJ....127..449M,Martin_inprep} is less restrictive and places an upper limit at about $\lesssim$100~Myr. The H$\alpha$ criterion of \citet{2000AJ....120..479A} was defined to select candidate in the range $\approx$5--50~Myr. The 21 outliers are therefore young ($\lesssim$100~Myr) free floating objects that could either have been ejected from neighboring associations, belong to unknown coincident foreground associations, or have recently experienced dynamical interactions within the Upper Scorpius association. Figure~\ref{map} shows a map of the Upper Scorpius region with the sample over-plotted. 

\subsection{Accretor frequency among outliers and members}
\begin{figure}
   \centering
   \includegraphics[width=0.45\textwidth]{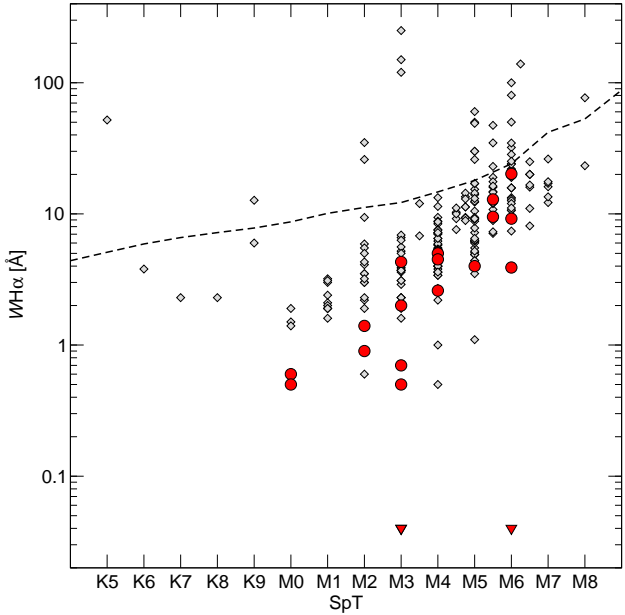}
   \caption{H$\alpha$ equivalent width (in~\AA) versus spectral type. Co-moving objects are represented with grey diamonds, and candidate outlier with red dots. Upper limits are represented with triangles and the same color code. The chromospheric criterion of \citet{2003AJ....126.2997B} is represented with a dashed line. According to this criterion, none of the outliers is accreting. }
              \label{accretion}%
    \end{figure}

Low resolution optical spectra have been obtained for all the objects, and measurements of H$\alpha$ equivalent width have been published. Our outlier selection criterion based on the proper motion membership probability is completely independent from the H$\alpha$ equivalent width, and hence we can use H$\alpha$ to make an unbiased comparison between the accretor frequency among kinematic outliers and members in the co-moving group. Figure~\ref{accretion} shows the H$\alpha$ equivalent width as a function of spectral type, and the chromospheric criterion defined by \citet{2003AJ....126.2997B} to distinguish between chromospheric activity and accretion induced H$\alpha$ emission. According to this criterion, a total of 25 of the 229 co-moving members found in the combined USNO-B and UCAC2 samples are accretors, while none of the 19 outlier candidates is accreting. The accretor frequency for members (11\%) is higher than {\bf the upper limit observed for} outliers ($<$5\%), but only at the 77\% confidence level, as determined from the two-tailed Fisher's Exact Test \citep{Fisher22JRSS}. 
The difference between the outliers and members accretor frequencies is not statistically conclusive.

\subsection{Disk frequency among outliers: mid-infrared observations}
\begin{figure}
   \centering
   \includegraphics[width=0.45\textwidth]{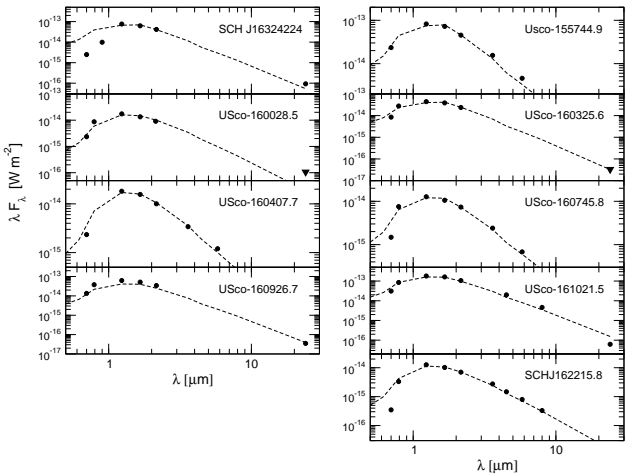}
   \caption{Spectral energy distributions of the 9 outliers with available {\it Spitzer} photometry (black dots). The NextGen theoretical spectra of the corresponding effective temperature are over-plotted as dashed lines. The single flux upper limit at 24~$\mu$m available for USco-160028.5 does not allow to conclude on the presence of mid-IR excess. None of the other nine outliers display mid-IR excess.}
              \label{seds}%
    \end{figure}

We searched the {\it Spitzer Space Telescope} public archive for images of the sample of 19 outliers. We found that 9 of them fall in the field of view of IRAC and/or MIPS observations as part of programs 148 (P.I. Meyer), 177 (P.I. Evans), 20069 (P.I. Carpenter) and 58 (P.I. Rieke). We retrieved the pipeline processed data and extracted PSF photometry of the outliers when detected, and flux upper limits when undetected. The photometry was extracted using standard PSF photometry procedures within the Interactive Data Language. Uncertainties were estimated from the Poisson noise weighted by the coverage maps of the mosaics, but are in general dominated by the flux calibration uncertainties. Upper limits were estimated by adding artificial stars of decreasing luminosity until the 3-$\sigma$ detection algorithm misses it. The results are given in Table~\ref{table_spitzer} and Fig.~\ref{seds}. The mid-infrared (mid-IR) observations allow us to put constraints on the presence of hot circumstellar inner disks around these outliers. The single flux upper limit at 24~$\mu$m available for USco-160028.5 does not allow us to conclude on the presence of mid-IR excess. None of the other nine outliers display mid-IR excess. Recent studies based on similar analyses found a mid-IR excess fraction of 22\%  among Upper Scorpius K0--M9 members \citep[37 out of 162 members for the combined surveys of ][]{2006ApJ...651L..49C,2007ApJ...660.1517S}, larger than that for outliers (0 out of 9) but at the 63\% confidence level only (as determined from the two-tailed Fisher's Exact Test). The difference is therefore statistically inconclusive.

\subsection{Photometric distances}
We use the spectral energy distribution (SED) to estimate the photometric distances to the outliers. The SED are unreddened using the extinction measurements provided by the different discoverers, or assuming an average A$_{\rm V}$=1.4~mag as reported in \citet{2002AJ....124..404P} when no measurement is available. The spectral types are translated into effective temperatures using the temperature scale given in \citet{2003ApJ...593.1093L}. Considering the age limit given in section~\ref{nature}, we derive the photometric distance by fitting the unreddened SED with the corresponding NextGen photospheric SED \citep{1998A&A...337..403B} for ages of 5, 50 and 100~Myr. In each case the best fit is obtained using a $\chi^{2}$ minimization with the distance being the only free parameter. The results are displayed in Table~\ref{table_dist}. Using Hipparcos, \citet{1999AJ....117..354D} estimated an average distance to the association of 145~pc, while \citet{1998AJ....115..351M} suggested the presence of some LMS members as close as 80~pc from the Sun. In order to be conservative, we hereafter consider that any object within 145$\pm$65~pc is in the distance range of the  association. If 5~Myr old, the outliers must be located approximately within 99--443~pc and 11 of them have photometric distances consistent with the association. The remaining eight all have distances greater than that of the association, and are therefore most likely background coindicences. If 100~Myr old, the outliers must be located approximately within 33--131, and 8 of them have photometric distances consistent with the boundaries of the association. 

\begin{center}
\begin{table}
\caption{{\it Spitzer} IRAC and MIPS photometry of the outliers}
\begin{scriptsize}
\label{table_spitzer}
\begin{tabular}{lccccc}\hline\hline
Objects          & 3.6~$\mu$m   & 4.5~$\mu$m   & 5.8~$\mu$m   & 8.0~$\mu$m   & 24~$\mu$m \\
                 & [mJy]        & [mJy]        & [mJy]        & [mJy]        &  [mJy]    \\
\hline
SCHJ16324224     & \nodata      & \nodata      & \nodata      & \nodata      & 0.74$\pm$0.14 \\
Usco-155744.9    & 18.6$\pm$1.9 & \nodata      & 8.9$\pm$0.9  & \nodata      & \nodata \\
USco-160028.5    & \nodata      & \nodata      & \nodata      & \nodata      & $<$0.9 \\ 
USco-160325.6    & \nodata      & \nodata      & \nodata      & \nodata      & $<$0.3 \\
USco-160407.7    & 4.1$\pm$0.4  & \nodata      & 2.3$\pm$0.2  & \nodata      & \nodata \\
USco-160745.8    & 2.9$\pm$0.3  & \nodata      & 1.3$\pm$0.2  & \nodata      & \nodata \\
USco-160926.7    & \nodata      & \nodata      & \nodata      & \nodata      & 0.3$\pm$0.03 \\
USco-161021.5    & \nodata      & 30.0$\pm$0.3 & \nodata      & 12.3$\pm$0.1 & 0.5$\pm$0.05 \\
SCHJ162215.8    & 3.3$\pm$0.3  & 2.2$\pm$0.2  & 1.6$\pm$0.2  & 0.9$\pm$0.1  & \nodata \\
\hline
\end{tabular}
\end{scriptsize}
\end{table}
\end{center}

\begin{center}
\begin{table}
\caption{Photometric distances to the outliers assuming various ages}
\begin{scriptsize}
\label{table_dist}
\begin{tabular}{lcccc}\hline\hline
  Objects               & SpT           & d (5~Myr) & d (100~Myr) \\
                        &               & [pc]      & [pc]        \\
\hline
SCHJ15582566-18260865 & M6 & 115 & 44 \\
SCHJ16084058-22255726 & M4 & 439 & 130 \\
SCHJ16221577-23134936 & M6 & 176 & 68 \\
SCHJ16324224-23165644 & M5.5 & 99 & 33 \\
USco-155744.89-222351.1 & M2 & 294 & 85 \\
USco-155848.44-224658.4 & M0 & 443 & 131 \\
USco-155930.21-225126.2 & M4 & 293 & 87 \\
USco-160017.34-221811.1 & M6 & 148 & 57 \\
USco-160028.47-220922.6 & M6 & 170 & 66 \\
USco-160054.45-224908.9 & M3 & 195 & 62 \\
USco-160142.6-222923 & M0 & 270 & 81 \\
USco-160236.21-191732.5 & M3 & 346 & 109 \\
USco-160325.6-194438 & M2 & 379 & 108 \\
USco-160407.73-194857.9 & M5 & 186 & 69 \\
USco-160522.69-205112.1 & M4 & 159 & 47 \\
USco-160745.74-203055.7 & M3 & 422 & 134 \\
USco-160845.6-182443 & M3 & 150 & 48 \\
USco-160926.71-192502.5 & M3 & 210 & 67 \\
USco-161021.5-194132 & M3 & 107 & 34 \\
\hline
\end{tabular}
\end{scriptsize}
\end{table}
\end{center}

\section{Origin of the outliers}
Proper motions only are not enough to trace back the origin of objects, as the age, distance and radial velocities would be required to fully understand where they come from. Over long timescales (typically greater than $\approx$10\,000~yr), the radial motion will cause a significant change in the distance and therefore in the apparent transverse velocity. Associations might also be rotating, adding another unknown to the equation. With these limitations in mind, we try to discuss the origin of the outlier candidates identified in our analysis.

\subsection{Ejected members of Upper Scorpius}
Using the current proper motions we can tentatively trace back the location of the outliers and check whether they could originate from the Upper Scorpius association. Outliers identified in our study have proper motion amplitudes in the range 25$\sim$230~mas yr$^{-1}$. For them to remain in the field of the discovery surveys ($\approx$15\degr$\times$15\degr at most, see Fig~\ref{map}), the interactions should have occurred less than 0.25--2.2~Myr ago (assuming a null radial velocity and constant proper motion). The probability for dynamical interactions to occur decreases rapidly with time after the first few 0.1~Myr of the formation of the parent proto-stellar cores \citep{2009MNRAS.392..590B}. Dynamical interactions within the loose Upper Scorpius association are therefore extremely unlikely within the past 2.2~Myr.  These simple statistical considerations suggest that the fastest outlier candidates are most likely not ejected members of the Upper Scorpius association itself.

\subsection{Upper Scorpius A and B}
Even though the original surveys used in our study were all performed in Upper Scorpius~A, some spatial overlap with Upper Scorpius~B (USco~B) cannot be ruled out. Some of the outliers candidates identified in our study could be either coincident USco~B members or ejected USco~B members. \citet{2007ApJ...662..413K} assigned a mean proper motion of (-10,-25)~mas yr$^{-1}$ for USco~A and (-21.3,-25.5)~mas yr$^{-1}$ for USco~B. The difference between the two subgroups mean proper motions is of the same order or smaller than the uncertainties on the USNO-B and UCAC2 measurements, so that an USco~B members would most likely fall within the selection criterion defined in section~\ref{membership}. While the age and distance of USco~A are well constrained thanks to the presence of high-mass members with {\it Hipparcos} measurements, the properties and history of USco~B are poorly known, making it difficult to test whether some of the outliers could be USco~B members.

\subsection{Neighboring associations}
Upper Scorpius is part of the Sco-Cen complex  \citep{1989A&A...216...44D} and is surrounded by the Upper Centaurus Lupus (13~Myr at 160~pc) and Lower Centaurus Crux (10~Myr at 118~pc) associations. Comparing the outliers proper motions to the mean proper motion of these loose associations would be meaningless given the distances between these associations and the corresponding large projection effects. Improved proper motion measurements as well as distance and radial velocity measurements will be required to perform a detailed comparison in the $(U,V,W)$ plane.

The 1~Myr old $\rho-$Oph association is located on the south-eastern part of Upper Scorpius (see Fig.~\ref{map}) and is sufficiently close that projection effects on the proper motions are smaller than the typical uncertainties of the USNO-B and UCAC2 measurements. \citet{2008AN....329...10M} recently estimated its distance to be 139$\pm$6~pc. Assuming that they are young, a number of outliers have photometric distances in the range 100--170~pc and could possibly originate from $\rho-$Oph. 

\subsection{Nearby coincident associations}
Even within their large uncertainties, the photometric distances indicate that a number of outliers are nearby and cannot belong to the more distant Upper Scorpius association. These objects could be members of coincident young nearby associations such as AB Dor or $\beta-$Pic \citep{2008arXiv0808.3362T}. Membership of these associations can only be inferred in the $(U,V,W)$ plane and therefore requires accurate radial velocity and distance measurements.

\section{Discussion}
Our study shows that 19 objects out of the combined sample of 251 objects (USNO-B+UCAC2) selected photometrically and confirmed spectroscopically have proper motion inconsistent with that of the Upper Scorpius association. If these outliers have the same age as the Upper Scorpius assocition, then 11 of them have photometric distances consistent with that of the association and could be genuine members having recently experienced dynamical interactions. Although this latter hypothesis cannot be rules out, the low probability for such interactions to have occured within the loose Upper Scorpius association nevertheless suggests that most of these outliers are background or foreground coincidences. Such contaminants are a well known problem that plagues the study of the initial mass function low mass end based on photometric surveys only. \citet{2008ApJ...689.1295K} and \citet{2009A&A...497..973V} have recently reported a number of young free floating very low mass objects in the field, suggesting that the contamination of photometrically selected samples of young associations could remain significant even after spectroscopic confirmation. 

With 19 outliers among a total sample of 251 objects, we derive an observed outlier frequency of 7.6$\pm$1.8\%. The current statistics has several limitations and this value should be considered with caution. As previously mentioned, our membership criterion was not meant to derive a statistically complete sample of outliers. Instead it was defined conservatively to separate the most obvious outliers from the association members. It therefore probably misses a number of contaminants. Moreover limiting the search radius to 3\arcsec\, additionally introduces a bias toward smaller proper motions, but increasing it also considerably increases the number of erroneous sources as objects with greater proper motions have a higher probability of being spurious entries in the catalogues. UCAC2 sources with proper motion greater than 0\farcs2 yr$^{-1}$ were systematically rejected from the catalogue.  The outlier frequency derived from the current analysis 
should be regarded as a preliminary lower limit.

\bibliographystyle{aa}

\section{Conclusions and future prospects}
We have cross-matched catalogues of low, very low mass and substellar objects classified as members of the Upper Scorpius association based on their photometric and spectroscopic characteristics with the USNO-B and UCAC2 catalogues. Our conclusions are threefold:
\begin{itemize}
\item a significant fraction of them have proper motion measurements reported in the USNO-B and UCAC2 catalogues. We find an observed frequency of 7.6$\pm$1.8 (19 out of 251) of objects with proper motions inconsistent with that of the co-moving group. This value is only a preliminary lower limit since our analysis suffers from a number of biases and incompleteness. Improved studies based on statistically complete samples and using more accurate proper motion measurements are required to derive accurate values of the outlier frequency. 
\item The outliers must be either free-floating young objects wandering after being ejected from a neighboring association, members of coincident foreground associations, or members of the association having experienced dynamical interactions within the past 0.2--2.2~Myr. The observed accretor and disc frequencies are lower among outliers than among members, but the current small-number statistics does not allow us to draw firm conclusions. 
\item to minimize this source of contamination, comprehensive surveys should be based on photometrically, spectroscopically and kinematically confirmed members. 
\end{itemize}

Follow-up observations of the non-members are required to understand their origin. In particular, the study of their multiplicity and how it compares to Upper Scorpius members could provide hints on whether these objects went through ejection events. Only short separation multiple systems are expected to survive the dynamical interactions responsible for the ejection. These processes are also expected to imprint a distinct secondary mass function on multiple systems. Finally, radial velocity and parallax measurements could help trace back their origin and whether they simply belong to unknown foreground associations or were ejected from the association or from a neighboring association. In the latter cases, kinematics offers a unique opportunity to identify ejected young low and very low mass objects. Understanding their origin and history holds important clues on the contribution of ejection phenomenons to the formation of low and very low mass objects. Kinematics will thus certainly play a key role to identify objects having experienced dynamical interactions, to quantify the contribution of ejection to the formation of very low mass objects, and understand the consequences of dynamical ejection on the formation and evolution of very low mass objects.

\begin{acknowledgements}
We are grateful to our anonymous referee for the prompt report, constructive criticism and important suggestions and corrections which helped improving tremendously the paper. H. Bouy acknowledges funding from the European Commission's Sixth Framework Program as a Marie Curie Outgoing International Fellow (MOIF-CT-2005-8389). E. L. Mart\'\i n acknowledges funding from the Spanish Ministerio de Educación y Ciencia through grant AyA2007-67458. 
This research has made use of the VO Theoretical data Server of the Spanish Virtual Observatory (SVO) operated by the Laboratory for Space Astrophysics and Theoretical Physics (LAEFF). This work has made use of the Vizier Service provided by the Centre de Donn\'ees Astronomiques de Strasbourg, France \citep{Vizier}. This research made use of Montage, funded by the National Aeronautics and Space Administration's Earth Science Technology Office, Computation Technologies Project, under Cooperative Agreement Number NCC5-626 between NASA and the California Institute of Technology. Montage is maintained by the NASA/IPAC Infrared Science Archive. This research has made use of the USNOFS Image and Catalogue Archive operated by the United States Naval Observatory, Flagstaff Station (http://www.nofs.navy.mil/data/fchpix/). This work made use of the USNO CCD Astrograph Catalog \citep[http://ad.usno.navy.mil/ucac/, ][]{2004AJ....127.3043Z}. This work has made use of the DENIS survey. The DENIS project has been partly funded by the SCIENCE and the HCM plans of the European Commission under grants CT920791 and CT940627. It is supported by INSU, MEN and CNRS in France, by the State of Baden-W\"urttemberg  in Germany, by DGICYT in Spain, by CNR in Italy, by FFwFBWF in Austria, by FAPESP in Brazil, by OTKA grants F-4239 and F-013990 in Hungary, and by the ESO C\&EE grant A-04-046. Jean Claude Renault from IAP was the Project manager.  Observations were  carried out thanks to the contribution of numerous students and young scientists from all involved institutes, under the supervision of  P. Fouqu\'e,  survey astronomer resident in Chile. This work is based in part on observations made with the Spitzer Space Telescope, which is operated by the Jet Propulsion Laboratory, California Institute of Technology under a contract with NASA. Support for this work was provided by NASA through an award issued by JPL/Caltech. 

\end{acknowledgements}

\begin{appendix}

\section{SCH~J16202523-23160347}

\begin{figure}
   \centering
   \includegraphics[width=0.45\textwidth]{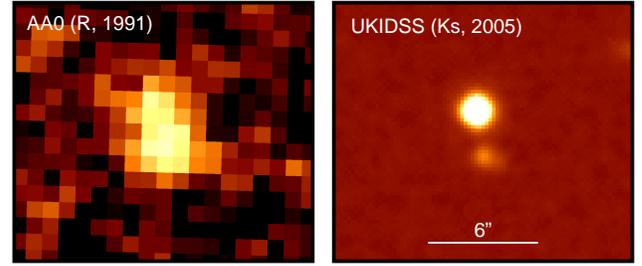}
   \caption{AAO-R (1991.23) and UKIDSS-Ks (2005.46) images of SCH~J16202523-23160347. The source is clearly elongated in the AAO image, and resolved as a triple system in the UKIDSS image. North is up and east is left, and the scale is indicated. }
              \label{schj162025}%
    \end{figure}

Figure~\ref{schj162025} shows that SCH~J16202523-23160347 is either resolved or elongated in most USNO-B images as well as in the 2MASS images. It is clearly resolved as a visual triple system in the H and K-band UKIDSS images. Unfortunetaly, the current UKIDSS data release does not include Z, Y or J band images of this field. Whereas at optical wavelengths the southern component is more luminous, the northern component is significantly brighter from 1.6~$\mu$m to longer wavelengths. We call the brightest optical component in the following the primary. The UKIDSS catalogue reports detection of only the primary (blended with the tertiary) and the secondary. In order to measure the astrometry and photometry of all three components, we retrieved the H and K band images from the UKIDSS server and extracted the photometry using PSF photometry with the \emph{Starfinder} code \citep{2000SPIE.4007..879D}. The zeropoints were computed using the UKIDSS photometry of component B. The results are given in Table~\ref{table_schj162025}

The quality of the USNO-B images does not allow to measure useful relative astrometry of the components to check common proper motion. Assuming that the system is member of Upper Scorpius and shares the mean proper motion of the association, a background source would have moved by $\approx$0\farcs6 between these 2 epochs. Such a motion is unfortunately smaller than the relative astrometric accuracy achievable in the USNO-B images.

\begin{figure}
   \centering
   \includegraphics[width=0.45\textwidth]{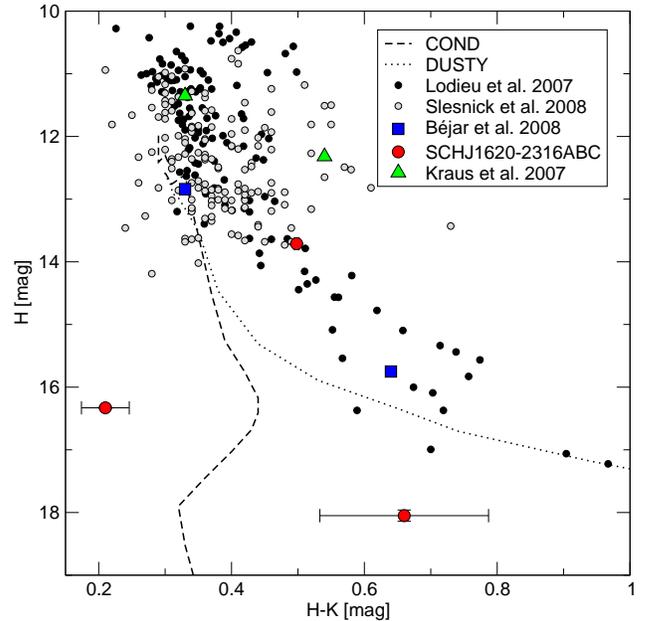}
   \caption{H vs H-K color-magnitude diagram for SCH~J16202523-23160347 and compared to Upper Scorpius very low mass members from the literature including 2 very low mass multiple systems \citep{2008ApJ...673L.185B,2007ApJ...664.1167K}. The B component has luminosities and colors consistent with those of very low mass members. The A et B components luminosities and colors are inconsistent with those of very low mass members. The DUSTY \citep{2002A&A...382..563B} and COND \citep{2003A&A...402..701B} isochrones at the distance of Upper Sco are over-plotted.}
              \label{h_hk}%
    \end{figure}

 The unresolved system was classified as M5.5 by \citet{2008ApJ...688..377S}, but their spectrum might have been affected by the dominating contribution of the blue primary. The H and K band photometry corresponds to a spectral type of $\approx$M8. SCH~J16202523-23160347A is therefore very likely a brown dwarf. Figure~\ref{h_hk} shows a color-magnitude diagram of known members from the literature with the three components over-plotted. While B has colors and luminosities consistent with those of very low mass members of the association, A and C lie far off the tracks. They are therefore most likely not members of the association but coincident sources.

\begin{center}
\begin{table}
\caption{Photometric and astrometric properties of the visual triple system SCH~J16202523-23160347}
\begin{scriptsize}
\label{table_schj162025}
\begin{tabular}{lcccc}\hline\hline
Component    & H                & K                & Sep.  &  P.A  \\
\hline
B            & 13.711$\pm$0.002 & 13.213$\pm$0.003 & 2\farcs618$\pm$0.003 & 8.68\degr$\pm$0.09\degr \\
A            & 16.33$\pm$0.02   & 16.12$\pm$0.03   &                      &                         \\
C            & 18.05$\pm$0.09   & 17.39$\pm$0.09   & 0\farcs915$\pm$0.015 & 245\degr$\pm$4\degr     \\
\hline
\end{tabular}
\end{scriptsize}
\end{table}
\end{center}

\end{appendix}

\Online
\newpage
\onecolumn

\onltab{3}{
\begin{deluxetable}{lcccccc|cccccccccc|cccc}
\tablecaption{Sample of Upper Scorpius objects with a counterpart in the USNO-B only or USNO-B and UCAC2 catalogues\label{table_sample}}
\tabletypesize{\scriptsize}
\rotate
\tablewidth{0pt}
\tablehead{
\multicolumn{7}{c|}{Literature} & \multicolumn{10}{c|}{USNO-B} & \multicolumn{4}{c}{UCAC2} \\
\hline
\colhead{Name} &  \colhead{R.A}     & \colhead{Dec.}    & \colhead{SpT} & \colhead{R} & \colhead{I} & \colhead{$W(\rm H\alpha)$} & \colhead{$r$}       & \colhead{Mean} & \colhead{N$_{\rm obs}$} & \colhead{R1}     & \colhead{R2} & \colhead{I}     & \colhead{$\mu_{\alpha} \cos \delta$} & \colhead{$\mu_{\delta}$} & \colhead{$\mu_{\rm Prob}$} & \colhead{Prob.} &\colhead{$r$}     & \colhead{$\mu_{\alpha} \cos \delta$} & \colhead{$\mu_{\delta}$}  & \colhead{Prob.} \\
\colhead{}     &  \colhead{(J2000)} & \colhead{(J2000)} & \colhead{}    & \colhead{[mag]} & \colhead{[mag]} & \colhead{[\AA]} & \colhead{[\arcsec]} & \colhead{Epoch} & \colhead{} & \colhead{[mag]}  & \colhead{[mag]} & \colhead{[mag]} & \colhead{[mas yr$^{-1}$]} & \colhead{[mas yr$^{-1}$]} & \colhead{[0.1]} & \colhead{[\%]} & \colhead{[\arcsec]} & \colhead{[mas yr$^{-1}$]} & \colhead{[mas yr$^{-1}$]} & \colhead{[\%]}
}
\startdata
\multicolumn{21}{c}{\citet{2002AJ....124..404P}} \\
\hline
USco-155625.64-224027.0 & 15 56 25.6 & -22 40 27.0 & M3 & 14.6 & \nodata & -4.2 & 0.812 & 1972.3 & 5 & 14.85 & 14.64 & 13.17 & -8$\pm$3 & -18$\pm$4 & 9 & 99.9 & 0.236 & -9.9$\pm$8.9 & -18.9$\pm$8.9 & 99.9       \\
USco-155629.51-225657.7 & 15 56 29.5 & -22 56 57.7 & M3 & 14.3 & \nodata & -3.1 & 0.680 & 1972.3 & 5 & 14.52 & 14.35 & 13.03 & -4$\pm$4 & -12$\pm$3 & 9 & 99.9 & \nodata & \nodata & \nodata & \nodata       \\
USco-155706.41-220605.7 & 15 57 06.4 & -22 06 05.7 & M4 & 16.2 & \nodata & -3.6 & 0.298 & 1972.3 & 5 & 16.44 & 16.25 & 14.30 & -18$\pm$6 & -6$\pm$6 & 9 & 85.9 & \nodata & \nodata & \nodata & \nodata       \\
USco-155728.49-221905.1 & 15 57 28.4 & -22 19 05.1 & M5 & 16.7 & \nodata & -6.2 & 1.085 & 1972.3 & 5 & 16.78 & 16.83 & 14.64 & -8$\pm$9 & -26$\pm$4 & 9 & 99.9 & \nodata & \nodata & \nodata & \nodata       \\
USco-155729.87-225843.7 & 15 57 29.8 & -22 58 43.7 & M4 & 15.8 & \nodata & -7.0 & 0.832 & 1972.3 & 5 & 15.95 & 15.80 & 14.02 & -6$\pm$0 & -16$\pm$7 & 8 & 99.9 & \nodata & \nodata & \nodata & \nodata       \\
USco-155742.50-222605.3 & 15 57 42.5 & -22 26 05.3 & M3 & 15.7 & \nodata & -2.9 & 0.263 & 1972.3 & 5 & 15.98 & 15.73 & 14.01 & -18$\pm$10 & -22$\pm$5 & 9 & 99.3 & \nodata & \nodata & \nodata & \nodata       \\
USco-155744.89-222351.1\tablenotemark{ a} & 15 57 44.8 & -22 23 51.1 & M2 & 14.3 & \nodata & -0.9 & 0.178 & 1972.3 & 5 & 14.75 & 14.11 & 13.29 & 2$\pm$11 & 28$\pm$3 & 9 & $\ll$0.1 & 0.617 & 17.8$\pm$8.8 & 10.4$\pm$9.0 & 0.1       \\
\enddata
\tablenotetext{ a}{Object classified as outlier ($\mathcal{P}<$0.1\% within the uncertainties, see Fig.~\ref{vectorpointdiag}).}
\tablenotetext{ b}{Resolved as a multiple system and discarded from the list of outlier candidates.}
\tablecomments{The R and I photometric systems used by the various authors are different. A direct comparison between the surveys is therefore not possible. $r$ is the distance between the object and its closest USNO-B/UCAC2 counterpart. $\mu_{\rm Prob}$. is the probability estimator for the likelihood that the proper motion is correct as defined in the USNO-B catalogue. This latter estimator ranges between 0.1 and 0.9, a larger value corresponding to a higher probability of being correct. N$_{\rm obs}$ is the number of epochs used for the proper motion measurement in the USNO-B catalogue. The membership probability are indicated in columns 17 and 21 for the USNO-B and UCAC2 catalogues respectively.}
\end{deluxetable}
}

\newpage
\cleardoublepage
\onltab{4}{
\begin{center}
\begin{deluxetable}{lccccccccc}
\tablecaption{Sample of Upper Scorpius objects with a counterpart in the UCAC2
catalogue only\label{table_ucac2_only}}
\tabletypesize{\scriptsize}
\tablewidth{0pt}
\tablehead{
\colhead{Name} &  \colhead{R.A}     & \colhead{Dec.}    & \colhead{SpT} & \colhead{R} & \colhead{$W(\rm H\alpha)$} & \colhead{$r$}  & \colhead{$\mu_{\alpha} \cos \delta$} & \colhead{$\mu_{\delta}$} & \colhead{Prob.} \\
\colhead{}     &  \colhead{(J2000)} & \colhead{(J2000)} & \colhead{}    & \colhead{[mag]} & \colhead{[\AA]} & \colhead{[\arcsec]} & \colhead{[mas yr$^{-1}$]} & \colhead{[mas yr$^{-1}$]} & \colhead{[\%]} 
}
\startdata
USco-155655.5-225839 & 15 56 55.5 & -22 58 39 & M0 & 13.1 & -1.9 & 1.533 & -11.8$\pm$4.6 & -21.5$\pm$4.6 & 99.9 \\
USco-160132.9-224231 & 16 01 32.9 & -22 42 31 & M0 & 14.6 & -1.4 & 0.957 & -9.9$\pm$4.6 & -8.5$\pm$4.6 & 99.5 \\
USco-160142.6-222923\tablenotemark{ a} & 16 01 42.6 & -22 29 23 & M0 & 13.5 & -0.6 & 1.332 & -50.1$\pm$8.8 & -54.2$\pm$8.8 & $\ll$0.1 \\
USco-160325.6-194438\tablenotemark{ a} & 16 03 25.6 & -19 44 38 & M2 & 15.4 & -1.4 & 0.578 & -49.7$\pm$8.8 & 33.4$\pm$8.9 & $\ll$0.1 \\ 
USco-160428.4-190441 & 16 04 28.4 & -19 04 41 & M3 & 13.5 & -5.0 & 0.567 & -3.1$\pm$8.7 & -17.2$\pm$8.8 & 99.8 \\
USco-160528.5-201037 & 16 05 28.5 & -20 10 37 & M1 & 14.1 & -1.9 & 0.915 & -6.9$\pm$8.7 & -24.0$\pm$8.6 & 99.9 \\
USco-160545.4-202308 & 16 05 45.4 & -20 23 08 & M2 & 14.4 & -35.0 & 1.033 & 9.5$\pm$9.6 & -2.4$\pm$9.1 & 77.9 \\
USco-160647.5-202232 & 16 06 47.5 & -20 22 32 & M2 & 13.9 & -3.2 & 0.212 & -29.7$\pm$8.8 & -10.7$\pm$8.8 & 95.4 \\
\enddata
\tablenotetext{ a}{Object classified as candidate interloper ($\mathcal{P}<$0.1\% within the uncertainties, see Fig.~\ref{vectorpointdiag}).}
\tablenotetext{ b}{Resolved as a {\bf visual} multiple system and discarded from the list of outlier candidates.}
\tablecomments{All the objects come from \citet{2002AJ....124..404P} sample. $r$ is the distance between the object and its closest UCAC2 counterpart. The membership probability is indicated in the last column.}
\end{deluxetable}
\end{center}
}

\end{document}